\documentstyle[pra,epsf,aps]{revtex}

  \newcommand{\dpp}[2]{\frac{\partial #1}{\partial #2}}

  \newcommand{\psibar}{{\overline{\psi}}}

  \newcommand{\sla}[1]{\sigma_{\mu} #1_{\mu}}

  \newcommand{\brpu}[1]{{\xi_#1}^+_\uparrow}
  \newcommand{\brmu}[1]{{\xi_#1}_\uparrow}	
  \newcommand{\brpd}[1]{{\xi_#1}^+_\downarrow}
  \newcommand{\brmd}[1]{{\xi_#1}_\downarrow}

  \newcommand{\bspu}{{\xi}^+_\uparrow}
  \newcommand{\bsmu}{{\xi}_\uparrow}	
  \newcommand{\bspd}{{\xi}^+_\downarrow}
  \newcommand{\bsmd}{{\xi}_\downarrow}

  \newcommand{\fspu}{{\psi}^+_\uparrow}
  \newcommand{\fsmu}{{\psi}_\uparrow}	
  \newcommand{\fspd}{{\psi}^+_\downarrow}
  \newcommand{\fsmd}{{\psi}_\downarrow}

  \newcommand{\mesb}{{\cal D}\xi{\cal D}\xi^*}
  \newcommand{\mesf}{{\cal D}\psi{\cal D}\psibar}

\title{Some spectral properties \\ of the one-dimensional disordered Dirac equation}
\author{\small Marc Bocquet}
\address{ Service de Physique Th\'eorique, C.E.A., Saclay.}
\date{\small 13 September 1998}

\draft
 
\begin{document}
\maketitle
 
\begin{abstract}
We study spectral properties of a one-dimensional Dirac equation with various disorder.
We use replicas to calculate the exact density of state and typical localization length
of a Dirac particle in several cases. We show that they can be calculated in any type
of disorder obeying a Gaussian white noise distribution. In particular,
we study the random electric potential model, as well as a mixed disorder case.
We also clarify the supersymmetric alternative derivation, even though it proves less
efficient than the replica treatment for such thermodynamic quantities.
We show that the smallest dynamical algebra in the Hamiltonian formalism is $u(1,1)$,
preferably to $u(n,n)$ in the replica derivation or $u(1,1|2)$ in the supersymmetric
alternative. 
Finally, we discuss symmetries in the disorder fields
and show that there exists a non trivial mapping between the
electric potential disorder and the magnetic (or mass) disorder.
\end{abstract}
 

 
\section{Introduction}

\label{sec1}

We plan to study some spectral properties of the one-dimensional Dirac Hamiltonian 
describing a Dirac particle of mass $m$ in a magnetic field  $\Phi$ and in an electric
(or scalar) potential $V$:
\begin{equation}
h=-i\sigma_z \partial_x+\Phi\sigma_x+M\sigma_y+V
\end{equation}
where the fields are random functions of the position $x$.

\subsection{A multiple disorder model}
We present a first motivation for this work.
In \cite{Lud} the authors drew what the space of coupling constants and the
renormalization group flow would look like for the 2D random Dirac Fermions.
Using bosonization and the renormalization group, they derived exact results
on localization/delocalization in random magnetic field, and in random mass field.
However, in the sole presence of an electric potential disorder,
the renormalization scheme flows away from the Gaussian fixed point to some
inaccessible quantum Hall-like transition fixed point.
What is so interesting with this type of disorder is that magnetic or/and mass disorder
do not suffice to account for the integer quantum Hall effect fixed point.
Thus the electric potential disorder seems to be a relevant field
in the flow to the true quantum Hall fixed point, even though there are just clues
this fixed point exists in the flow diagram of the 2D Dirac Fermions.
Chalker and Ho in \cite{Ho} also emphasized the importance of the electric potential 
disorder.
They took the continuum isotropic limit of the network model, known to capture the
key features of the localization/delocalization transition in the integer quantum Hall
effect \cite{Cha}.
They managed to map this continuum limit to the model of 2D Dirac Fermions in random fields.
In particular they showed there exists an electric potential random field
locally measuring the magnetic flux per plaquette in the network.

Though it should not shed much light on the 2D problem, we thought it would be
interesting to
solve thoroughly the 1D problem. In particular, we shall see that there exists a non 
trivial exact mapping between the magnetic disorder and the electric disorder. This is to be
compared to an argument proposed in \cite{Lud} relating the random mass field and the random
electric potential field.

\subsection{Localization}

A second motivation for this work is to know whether a true one-dimensional Dirac 
particle may be in delocalized states. It is already known that there is a zero energy
delocalized state in random mass field (or alternatively in random magnetic field).
Is there any other delocalized states when different types of disorder are present?
As we shall see, the answer is essentially no.

\subsection{From 2D Dirac equation with quasi-unidimensional disorder
to the one-dimensional Dirac equation}

The third motivation comes from the study of spectral properties of the quasi-unidimensional
2D Hamiltonian, describing a massive Dirac particle in a potential
vector and in an electric potential
\begin{equation}
h=-i\sla{\partial}-\sla{A}-M\sigma_z+V.
\end{equation}
The $\sigma_\mu$ ($\mu=x,y,z$) are the usual Pauli matrices.
The potential vector $\vec{A}$, the mass $M$, and the electric potential $V$
are all random functions obeying independent Gaussian laws.
The three random fields are supposed to depend only on $x$.

Because of gauge invariance the component $A_x$ can be chosen to be zero.
Fourier transforming on the y coordinate (and thus acting on wavefunctions
of the form $\psi(x,y)=\exp(-iky)\psi(x))$, one gets

\begin{equation}
h=-i\sigma_x \partial_x-(k+\Phi)\sigma_y-M\sigma_z+V
\end{equation}
where $\Phi$ stands for $A_y(x)$.
  
We now fix the average value of $M$ to be $m$ and set $k$ to be
the average value of $\Phi$ thus redefining it into a random function with
vanishing mean. The average value of $V$ can be chosen
to be zero up to rescaling the energy of the Hamiltonian.
By means of a unitary transformation which applies $\sigma_x$ to $\sigma_z$,
$\sigma_y$ to $-\sigma_x$, and $\sigma_z$ to $-\sigma_y$, and for convenience,
we shall adopt the following equivalent form of the Hamiltonian
\begin{equation}
h=-i\sigma_z \partial_x+\Phi\sigma_x+M\sigma_y+V
\end{equation}
with the given average values $\langle \Phi \rangle=\phi$, $\langle M \rangle=m$
and $\langle V \rangle=0$.
The fields $\Phi$, $M$ and $V$ will be (somehow arbitrarily) respectively called
{\em magnetic field}, {\em mass field} and {\em electric potential field}.
$h$ is precisely the Hamiltonian we would like to study.
Note that the random mass model (i.e. $\Phi=V=0$) which stems from a particular
form of $h$ has already been studied in \cite{Ovc} and \cite{Bou}. \\

The remainder of this article is organized as follows.
In sections \ref{sec1}, \ref{sec2} and \ref{sec3}, we map the resolvent
calculation to a quantum mechanical problem, using the replica trick.
In section \ref{sec4}, we derive exact formulas for the resolvent.
In section \ref{sec5}, we calculate the exact density of states and the typical
localization length in various configurations of disorder.
Then, in section \ref{sec6},
we sketch an alternative derivation using supersymmetry in a straightforward fashion.
Finally in section \ref{sec7} we study mappings between different disordered Hamiltonians.

	\section{Mapping to quantum mechanics}
\label{sec2}
We wish to calculate the resolvent of the Hamiltonian $h$ in order to extract
its density of states and its localization length.
In order to do so we will map the associated one-dimensional statistical field theory
onto a quantum mechanical problem using Feynman procedure.

	\subsection{Path integral representation of the resolvent}
The two-point resolvent is $G(x,y,\epsilon+i\eta)=
\mbox{tr}_{\mbox{\tiny spin}}G_{\alpha,\beta}(x,y,\epsilon+i\eta)$, where 
$\alpha$ and $\beta$ are spin indices. The spin components of the resolvent
are
\begin{equation}
G_{\alpha,\beta}(x,y,\epsilon+i\eta)=
\langle x,\alpha|\frac{1}{h-(\epsilon+i\eta)}|y ,\beta \rangle
\end{equation}
where $\eta$ is strictly positive.

Taking advantage of the translational invariance of the ensemble average
of the resolvent $G(\epsilon+i\eta)=\langle G(x,x,\epsilon+i\eta) \rangle$,
one can write
\begin{equation}
G(\epsilon+i\eta)=\lim_{L\rightarrow \infty} \frac{1}{2L} \int_{-L}^{L} dx
\langle \mbox{tr}_{\mbox{\tiny spin}}G_{\alpha,\beta}(x,y,\epsilon+i\eta) \rangle
\end{equation}
which also define $G(\epsilon+i\eta)$ as the {\em average} of the previous resolvent
$G(x,x,\epsilon+i\eta)$.
The integrand can then be formally represented by means of a Bosonic field
$\xi$ with two spinorial components $(\xi_\uparrow,\xi_\downarrow)$, so that
\begin{equation}
G(\epsilon+i\eta)=\lim_{L\rightarrow \infty} \frac{1}{2iL} \int_{-L}^{L} dx
\langle \mbox{tr}_{\mbox{\tiny spin}} \frac{\int \mesb \xi_\alpha^*(x)
\xi_\beta(x) \exp{(i\int_{-L}^{L}dxdy \xi^*(x) 
\langle x| h-(\epsilon+i\eta)|y \rangle \xi(y)})}
{\int \mesb \exp{(i\int_{-L}^{L}dxdy \xi^*(x) 
\langle x| h-(\epsilon+i\eta)|y \rangle \xi(y)})} \rangle.
\end{equation}
We then obtain
\begin{equation}
G(\epsilon+i\eta)=\lim_{L\rightarrow \infty} \frac{1}{2L}\dpp{}{\epsilon}
\langle \ln \left( \int \mesb \exp{(i\int_{-L}^{L}dx \xi^*(x) 
\langle x| h-(\epsilon+i\eta)|x \rangle \xi(x)}) \right) \rangle .
\end{equation}
The logarithm can be averaged using the replica trick. So replicating
$n$ times the two-component bosonic field, we are led to
\begin{equation}
G(\epsilon+i\eta)=\lim_{L\rightarrow \infty} \frac{1}{2L}\dpp{}{\epsilon}
\lim_{n\rightarrow 0}\dpp{}{n}\langle\int \mesb \exp{(i\int_{-L}^{L}dx
\sum_{k=1}^{n} \xi_k^*(x)
\langle x| h-(\epsilon+i\eta)|x \rangle \xi_k(x)}) \rangle.
\end{equation}

	\subsection{Imaginary time path integral}

Feynman formalism can be applied to map this one-dimensional path integral
(statistical field theory) to a 0-dimensional quantum field theory,
that is to say a mere quantum theory. We shall demonstrate that the corresponding
physical Hamiltonian is Hermitian in some precise sense. Here physical means
retaining $\eta=0$ in the Hamiltonian since the imaginary part of the energy
is a mathematical artifice to probe the analytic properties of the resolvent. 

Nevertheless when we re-establish $\eta \neq 0$ in order to obtain analytic properties
of the resolvent, we shall rewrite the partition function as one of a statistical
field theory in imaginary time:
\begin{equation}
\label{04}
Z_L(n)=\langle\int \mesb \exp{(-\int_{-L}^{L}dx
\sum_{k=1}^{n} \xi_k^*(x) 
\langle x| -ih+(-i\epsilon+\eta)|x \rangle \xi_k(x)}) \rangle
\end{equation}

We can now rename $-i\epsilon+\eta$ as $\epsilon$ and make it strictly positive
(then shrinking the previous $-i\epsilon$ part of the energy term to zero).
When all calculations are done, we will continue on $\epsilon$
to re-establish it to the imaginary axis.
From now on, $G(\epsilon)$ stands for $G(\epsilon+i\eta)$ after having continued
$-i\epsilon+\eta$ to the imaginary axis.

As a consequence, we have
\begin{equation}
\label{10}
G(\epsilon)=-i\lim_{L\rightarrow \infty} \frac{1}{2L}\dpp{}{\epsilon}
\lim_{n\rightarrow 0}\dpp{}{n}Z_L(n).
\end{equation}
The $i=\sqrt{-1}$ is placed here to compensate the fact that the first derivative
is now done with respect to the imaginary energy.

\subsection{Particle-hole transformation}

Actually the previous path integral formulas are ill defined for the reason
that they correspond to an unbounded quantum theory. To see it, we follow the
Feynman procedure, write the classical Hamiltonian associated to the
previous path-integrals, then quantize it.
The momenta for the spinorial fields $\xi_k$ are
\begin{equation}
\pi_{\xi_k}=\frac{\delta {\cal L}}{\delta \dot{\xi_k}}=- \xi_k^*\sigma_z
\end{equation}
where ${\cal L}$ is the Lagrangian density and the dot indicates derivative with respect to $x$.
So the Euclidean coherent state path integral formalism yields the
classical Hamiltonian
\begin{equation}
H_c=\sum_{k=1}^{n}\int dx \Pi_{\xi_k}.\partial_x {\xi_k}
+{\cal L}(\Pi_{\xi_k},{\xi_k})
\end{equation}
before any ensemble averaging.
To quantize this Hamiltonian, we impose the following
canonical commutators:
\begin{equation}
  [\xi_k(y), \Pi_{\xi_l}(x) ]=\delta_{k,l}\delta(x-y).
\end{equation}
This leads to a $\epsilon .\Pi_{\xi_k}\sigma_z \xi_k$ term in the quantum Hamiltonian
which makes it unbounded. This is due to the minus sign appearing within
the spin down components product.
To quantize the Hamiltonian safely we have the freedom
to make any transformation on the canonical operators whenever the
canonical commutation relations remain unmodified.
For Dirac Fermions the correct procedure would be to fill the Dirac sea
upon making a particle hole transformation. However, such a change of vacuum is
not appropriate for the Bosons that would condensate anyway.
So healing the unbounded Bosonic Hamiltonian
requires a stronger transformation. One can set
\begin{equation}
\Pi_{\xi_k}=({\xi_k}^+_\uparrow,-{\xi_k}_\downarrow) \qquad
\xi_k=({\xi_k}_\uparrow,{\xi_k}^+_\downarrow) .
\end{equation} 
The minus sign is due to heal the unboundedness of the spectrum.
The additional particle-hole transformation restores the commutation
relations destroyed by the inversion of the spin component operators.
The tribute to such a regularizing transformation is that the spinor momentum
$\Pi_{\xi_k}$ is not any more the Hermitian conjugate of $\xi_k$.
We shall see that the internal dynamical symmetry of the replicated
Hamiltonian is not $U(2n)$ any more but the non compact $U(n,n)$ group. 
We will call dynamical algebra (or improperly group), any algebra which 
generators are the sole building blocks of the Hamiltonian.

Let us define the Hermitian spin operators $\vec{\bf J}_{kl}=\frac{1}{2}\Pi_{\xi_k}
(i\sigma_x,i\sigma_y,\sigma_z)\xi_l$.
With the operators $Q_{kl}=\frac{1}{2}\Pi_{\xi_k}\xi_l$, they form the $4n^2$
generators of $u(n,n)$. In components, the spin operators read:
\begin{equation}
{\bf J}^x_{kl}=\frac{i}{2}(\brpu{k}\brpd{l}-\brmu{k}\brmd{l}) \qquad
{\bf J}^y_{kl}=\frac{1}{2}(\brpu{k}\brpd{l}+\brmu{k}\brmd{l}) \qquad
{\bf J}^z_{kl}= \frac{1}{2}(\brpu{k}\brmu{l}+\brpd{k}\brmd{l}) .
\end{equation}
We will notice that, be it before of after averaging,
 the Hamiltonian (or the Hamiltonian density) does only depend on the sum operators
\begin{equation}
Z^x=\sum_{k=1}^{n} {\bf J}^x_{kk} \qquad Z^y=\sum_{k=1}^{n} {\bf J}^y_{kk} 
\qquad Z^z=\sum_{k=1}^{n} {\bf J}^z_{kk} .
\end{equation}
Then we define the corresponding ladder operators, which we shall later use
 to generate physical sectors of the effective Hamiltonian
\begin{equation}
Z^+=Z^y-iZ^x \qquad Z^-=Z^y+iZ^x \qquad Z^0=Z^z .
\end{equation}
They obey $su(1,1)$ commutation rules
\begin{equation}
[Z^+,Z^-]=-2Z^0 \qquad [Z^0,Z^-]=Z^+ \qquad [Z^0,Z^-]=-Z^-
\end{equation}
which have to be completed with the identity $(Z^+)^\dagger=Z^-$, which tells us
which real form of $sl(2,C)$ the spin operators $\vec{Z}$ are the generators of.

This statement implies that the dynamical algebra can be reduced from $u(n,n)$
to $u(1,1)$. Such a strong simplification, which was not previously known,
also occur in the supersymmetric version of the derivation.

	\subsection{Averaging}

The disorder fields can be written away in the vector field $\Delta=(M,\Phi,V)$
with mean value $\Delta_0=(m,\phi,0)$.
The most general Gaussian disorder distribution
can be written using a covariance matrix $C$. $C$ is assumed to be symmetric
and positive definite, as it seems physically reasonable.
The Gaussian measure is then
\begin{equation}
P(\Delta)=\exp\left( -\frac{1}{2}\int_{-\infty}^{\infty} 
(\Delta-\Delta_0)^\dagger C (\Delta-\Delta_0) d\Delta \right).
\end{equation}
Performing the average, one obtains an effective Hamiltonian of the form
\begin{equation}
\frac{1}{2}H=-\Delta_0 . \vec{Z}-\vec{Z}^\dagger C^{-1}\vec{Z} 
\end{equation}
which does not depend any more on $x$.
We next rescale our Hamiltonian in order to eliminate the inconvenient $\frac{1}{2}$ 
factor.
Choosing to discard off-diagonal mixing terms in the covariance matrix,
 we will adopt the definitive form of our effective Hamiltonian:
\begin{equation}
H=\epsilon Z^0-imZ^x+i\phi Z^y+g_x (Z^x)^2+g_y (Z^y)^2+g_z (Z^z)^2
\end{equation}
where $g_x$, $g_y$ and $g_z$ denote the disorder strengths.
As announced, the effective Hamiltonian does only depend on the spin operators 
$\vec{Z}$.
From the definition we retained for a dynamical algebra, one immediately sees
that $u(1,1)$ is a dynamical algebra of $H$. 

	\section{Groundstate sector}
We shall now describe some  groundstate sector properties in a view to evaluate
$Z_L(n)$ when $L$ goes to infinity and $n$ goes to $0$.
\label{sec3}
	\subsection{Basis description}
One can look for the groundstate of the effective (averaged) Hamiltonian in the lowest
weight space generated by the dynamical symmetry $u(1,1)$.
Since the trace operator of $u(1,1)$ commutes with $H$, we can restrict
our dynamical symmetry to $su(1,1)$ in any labeled sector of the Hamiltonian.
The lowest weight of the desired space is equal to $n/2$
as the sum of the n spin-$1/2$ discrete lowest weight
representations of each individual spin. Call $|\Omega \rangle$ the lowest vector of
this particular representation. The states of this representation space,
generated by the action of operator $Z^+$ on $|\Omega \rangle$ are
$|k\rangle=\frac{(Z^+)^k}{k!} |\Omega \rangle$.
The norms of such states can be quite easily computed. We obtain
\begin{equation}
\langle k|k\rangle=\frac{\Gamma(k+n)}{\Gamma(n)\Gamma(k+1)}.
\end{equation}
Reducing the dynamical symmetry allows us to generate
the whole representation space as above.
Another advantage is that one can as easily calculate the action
of the spin operators in the $\{| k\rangle \}$ basis.
\begin{equation}
Z^0|k\rangle=(k+\frac{n}{2})|k\rangle \qquad
Z^+|k\rangle=(k+1)|k+1\rangle \qquad Z^-|k\rangle=(k+n-1)|k-1\rangle .
\end{equation}

	\subsection{Interpretation in terms of representation}

We shall see now that the replica limit $n$ goes to $0$ is mathematically
well defined.
As long as $n$ is a positive integer, the groundstate sector remains a lowest
weight representation space of $su(1,1)$ belonging to the discrete series.
When $n$ tends to 0 in the interval $]0,1[$, then the representation
space does not any more belong to  the discrete series but to the supplementary
series. In Vilenkin's notation, this representation is labeled by 
$(l=n-1,\epsilon=0)$ \cite{Vil}. The series of $|k\rangle$ states is not any more
truncated below by the action of $Z^-$ on $|1-n\rangle$.
Theoretically it spans down to $-\infty$.
Remarkably, when $n$ goes to $0$, we have 
\begin{equation}
\langle k|k\rangle=
\frac{\Gamma(k+n)}{\Gamma(n)\Gamma(k+1)}\longrightarrow \delta_{k,0}.
\end{equation}
So that, at the precise point corresponding to the physical disordered system ($n=0$),
the restriction of the scalar product within the groundstate sector degenerates.
The $|k\rangle$ states are null states except when $k=0$.
When $n$ is in $]0,\infty[$, one can define a proper scalar product in the groundstate 
sector for wavefunctions of the form $|\phi\rangle=\sum_{k=0}^{\infty} \phi_k |k\rangle$:
\begin{equation}
\langle \psi|\phi\rangle=\frac{1}{\Gamma(n)}\sum_{k=0}^{\infty}
\frac{\Gamma(k+n)}{\Gamma(k+1)} \psibar_k\phi_k
\end{equation}
which is identical to Vilenkin's formula (see also \cite{Vil}) but truncated below $k=0$.

	\section{Groundstate energy of the replicated Hamiltonian}
\label{sec4}
	\subsection{Groundstate wavefunction equation}
Recall the effective replicated Hamiltonian is
\begin{equation}
H=\epsilon Z^0-imZ^x+i\phi Z^y+g_x (Z^x)^2+g_y (Z^y)^2+g_z (Z^z)^2 .
\end{equation}
The expansion of the groundstate on the $\{ |k\rangle\}$ basis is
$|\Psi\rangle=\sum_{k=0}^{\infty} \zeta_k |k \rangle$.
It satisfies the eigenvalue equation $H|\Psi\rangle=\lambda(n)|\Psi\rangle$.
$\lambda(n)$ is expected to behave like $\lambda n +O(n^2)$ when $n$ goes to $0$.
Re-written in $\{ |k\rangle\}$ the basis, this eigenvalue equation yields 
\begin{eqnarray*}
&&(\epsilon(k+\frac{n}{2})+g_z(k+\frac{n}{2})^2)\zeta_k
-\frac{m}{2}(k \zeta_{k-1}-(k+n) \zeta_{k+1})
+i\frac{\phi}{2}(k\zeta_{k-1}+(k+n)\zeta_{k+1}) \\
&&+\frac{1}{4}k(k-1)(g_y-g_x)\zeta_{k-2}
+\frac{1}{2}(k(k+n)+\frac{n}{2})(g_y+g_x)\zeta_k
+\frac{1}{4}k(k+1)(g_y-g_x)\zeta_{k+2}=\lambda(n)\zeta_k
\end{eqnarray*}
The equation is valid for any integer $k\geq 0$.

	\subsection{Left and right eigenstates of the Hamiltonian}

For technical reasons, that is when $\eta\neq 0$, we must consider a non-Hermitian Hamiltonian.
As a consequence, Feynman's formula which gives probability amplitudes
of a quantum mechanical theory in terms of one-dimensional statistical
path integrals, must be written in terms of left and right eigenstates.
This follows from the resolution of unity used to derive the Feynman equivalence.
Indeed, the resolution of the identity is expressed in terms of right and left eigenstates.

A simplifying feature is that our Hamiltonian and its Hermitian conjugate are
unitarily related by means of the simple unitary transformation
$R=\exp(i\pi \sum_{k=1}^n {\bf J}^z_{kk})=\exp(i\pi Z^z)$. 
Left eigenstates lie in the dual vector space of the current Fock space.
In principle, there is no need to project them to the original vector space.
But doing so is convenient.
Using the transformation $R$, we can map those left eigenstates back into the Fock 
space:
\begin{equation}
|\psi\rangle_l=R|\psi\rangle_r.
\end{equation}

	\subsection{Groundstate wavefunction}

In order to apply Feynman formalism, we must normalize our groundstate wavefunction.
When $n\neq 0$, we have
\begin{equation}
\langle \Psi |_l \Psi\rangle_r=\frac{1}{\Gamma(n)}\sum_{k=0}^{\infty}
 (-1)^k \frac{\Gamma(k+n)}{\Gamma(k+1)} |\zeta_k|^2 .
\end{equation}
In the limit $n$ goes to zero, we obtain $||\Psi||^2=|\zeta_0|^2$.
So that the normalization condition can be chosen $\zeta_0(n=0)=1$.

\subsection{Hierarchical equations}

Since we are interested in the limit $n$ goes to $0$, we expand
the groundstate wavefunction and its eigenvalue in powers of $n$ (see \cite{Itz} for
the use of such scheme on the Halperin model).
At order $1$, we set the notations
\begin{equation}
\left\{
\begin{array}{l}
\lambda(n)=n\lambda+O(n^2) \\
\zeta_k(n)=\psi_k+n\chi_k+O(n^2)
\end{array}
\right.
\end{equation}
At order $0$, we find the equation for the groundstate of the physical
system (i.e. when $n=0$)
\begin{eqnarray}
\label{02}
&&(\epsilon+(g+\frac{g_x+g_y}{2})k)\psi_k-\frac{m}{2}(\psi_{k-1}-\psi_{k+1})
+i\frac{\phi}{2}(\psi_{k-1}-\psi_{k+1}) \nonumber \\
&&+\frac{1}{4}(g_x-g_y)(k-1)\psi_{k-2}
+\frac{1}{4}(g_x-g_y)(k+1)\psi_{k+2}=0 .
\end{eqnarray}

The result is an extension of an equation obtained by Luck \cite{Luc},
then Balents and Fisher \cite{Bal}.
The coefficients $\psi_k$ should bear a physical interpretation in terms of
amplitudes of multiple scattering on impurities.
At least formally, in the random mass model with zero mass average,
the coefficient $\psi_{2k}$ may be interpreted as the right-hand-side Berezinskii block $R_k$
(see \cite{Ber},\cite{Fab}).
The recursion relation on the Berezinskii blocks $R_k$ or on the
groundstate coefficients $\psi_{2k}$ are indeed the same
(the odd $\psi_{2n+1}$ do not contribute).
Similarly the coefficients of the left groundstate may be
interpreted as the left-hand-side Berezinskii block $\tilde{R}_k$.
Those blocks are the sum of all contributions due to scattering events happening on the right
and on the left of the point where the one-point Green function is computed.

Coming back to the hierarchical system, order $1$ yields the equation for $\chi_k$:
\begin{eqnarray*}
&& \left[ (\epsilon+(g+\frac{g_x+g_y}{2})k)\chi_k
-\frac{m}{2}(\chi_{k-1}-\chi_{k+1})
+i\frac{\phi}{2}(\chi_{k-1}-\chi_{k+1})  \right. \\
&& \left.+\frac{1}{4}(g_x-g_y)(k-1)\chi_{k-2}
+\frac{1}{4}(g_x-g_y)(k+1)\chi_{k+2} \right] k \\
&&+\frac{\epsilon}{2}\psi_k
+\frac{g_x+g_y}{2}(k+\frac{1}{2})\psi_k+g k \psi_k
+\frac{1}{4}(2k+1)(g_y-g_x)\psi_{k+2}+\frac{1}{2}(m+i\phi)\psi_{k+1}
=\lambda \psi_k .
\end{eqnarray*}

The equation at order $1$ can then be legitimately evaluated at $k=0$ and yields
the surprisingly simple result:
\begin{equation}
\label{08}
\lambda=\frac{\epsilon}{2}+\frac{g_x+g_y}{4}
+\frac{m+i\phi}{2}\frac{\psi_1}{\psi_0}
+\frac{g_y-g_x}{4}\frac{\psi_2}{\psi_0} .
\end{equation}

\subsection{Resolvent formula}
From formula (\ref{04}) it is sufficient to evaluate the partition function
$Z_L(n)$ in the limit where $L$ goes to infinity and the replica index $n$ goes to zero
to calculate the imaginary time resolvent. In that limit, we have
\begin{equation}
Z_L(n)\sim\langle \Psi|_l\exp(-H)|\Psi\rangle_r\sim\exp(-2L\lambda n+O(n^2)),
\end{equation}
so that with equation (\ref{10}) 
\begin{equation}
G(\epsilon)=2i\dpp{\lambda}{\epsilon}.
\end{equation}
Note that the $2$ factor is due to the adopted scaling convention of the Hamiltonian,
and consequently of $\lambda$.

\subsection{Sum formula}
	
In the previous subsection, we derived an efficient way to obtain the resolvent.
Now, we shall show that though it is more cumbersome, we can also write the 
resolvent as a sum as was done in \cite{Bal}.
At first, let us replicate the correlation function $n$ times without changing its value:
\begin{equation}
G(\epsilon)=\frac{1}{i}\langle \mbox{tr}_{\mbox{\tiny spin}}
\frac{\int \mesb \frac{1}{n}\sum_{k=1}^n {\xi_\alpha^k}^*(x) \xi_\beta^k (x)
 \exp{(-\int_{-\infty}^{\infty}dx
\sum_{k=1}^{n} \xi_k^*(x) \langle x| -ih+(-i\epsilon+\eta)|x \rangle \xi_k(x)})}
{\int \mesb \exp{(-\int_{-\infty}^{\infty}dx
\sum_{k=1}^{n} \xi_k^*(x) \langle x| -ih+(-i\epsilon+\eta)|x \rangle \xi_k(x)})}
 \rangle.
\end{equation}
In the limit where $n$ goes to $0$, the partition function denominator goes to $1$,
so that averaging becomes possible
\begin{equation}
G(\epsilon)=\frac{1}{i} \lim_{n\rightarrow 0}\langle \mbox{tr}_{\mbox{\tiny spin}}
\int \mesb \frac{1}{n} \sum_{k=1}^n {\xi_\alpha^k}^*(x) \xi_\beta^k (x)
\exp{(-\int_{-\infty}^{\infty}dx
\sum_{k=1}^{n} \xi_k^*(x) \langle x| -ih+(-i\epsilon+\eta)|x \rangle \xi_k(x)})
\rangle.
\end{equation}
After regularizing this correlator by particle-hole transformations as was done above,
and mapping the path integral to the equivalent quantum problem, we get
\begin{equation}
\label{11}
G(\epsilon)=-2i \lim_{n\rightarrow 0} \frac{1}{n}
\langle \Psi|_l Z^0 |\Psi\rangle_r.
\end{equation}
To render explicit this amplitude, we make use of our scalar product
\begin{eqnarray}
G(\epsilon)=-2i \lim_{n\rightarrow 0} \frac{1}{n}
\sum_{k=0}^\infty (-1)^k (k+\frac{n}{2})
|\psi_k|^2 \frac{\Gamma(k+n)}{\Gamma(n)\Gamma(k+1)}.
\end{eqnarray}
Eventually taking the limit $n\rightarrow 0$ we obtain:

\begin{eqnarray}
\label{05}
G(\epsilon)=-2i\sum_{k=0}^\infty (-1)^k |\psi_k|^2 .
\end{eqnarray}

	\section{Densities of states}
\label{sec5}
We now contemplate various cases of disorder and calculate the subsequent resolvents.

	\subsection{Random mass model}

The random mass model, which corresponds to the original Hamiltonian
$ h=-i\sigma_z \partial_x+M\sigma_y$ is the only previously (well-) known case.
$M$ is the random mass field and
follows the Gaussian measure $P(M)=\exp(-\frac{1}{2g}\int_{-\infty}^{\infty} dx (M-
m)^2)$.
It is equivalent to many one-dimensional physical models (see \cite{Tex} for a review on the
quantum supersymmetric Hamiltonian $h^2$).

After averaging the effective Hamiltonian one obtains
\begin{equation}
H=\epsilon Z^0-imZ^x+g (Z^x)^2.
\end{equation}
Correspondingly, its groundstate equation reads for $k\geq 1$
\begin{equation}
\label{01}
(\epsilon+\frac{g}{2}k)\psi_k-\frac{m}{2}(\psi_{k-1}-\psi_{k+1})
+\frac{1}{4}g(k-1)\psi_{k-2}+\frac{1}{4}g(k+1)\psi_{k+2}=0
\end{equation}
whereas the groundstate energy (\ref{08}) is slightly reduced to
\begin{equation}
\lambda=\frac{\epsilon}{2}+\frac{g}{4}
+\frac{m}{2}\frac{\psi_1}{\psi_0}
-\frac{g}{4}\frac{\psi_2}{\psi_0}.
\end{equation}

Such a	difference equation can be solved exactly.
In literature the technique of the generating
functional is often used to solve, when possible, quasi-linear recurrence equations.
We prefer to use Laplace method which, though
equivalent, is much more convenient in such circumstances.
Indeed it allows to pick up a particular solution of the equation
thanks to the choice of the appropriate contour in the complex
plane. This is all the more useful since we expect only one physically
acceptable solution among a finite dimensional vector space of solutions.

Let us detail how we use it.
We start giving an integral representation of $\psi_k$ in the complex plane
along a contour $\cal C$:
\begin{equation}
\label{12}
\psi_k=\int_{\cal C} dw K(k,w) \psi(w) .
\end{equation}

The kernel $K$  has to be chosen in order to yield the simplest differential equation
on $\psi(w)$. The contour $\cal C$ must be chosen so as to obtain
a solution different from zero. It also has to be chosen in a view to pick up
the desired solution in case there are more than one.
If the contour possesses boundaries, we require $\psi_w$
to tend to zero fast enough (depending on $K$) in their vicinities.

For our purpose, we choose the Melin kernel, $K(k,w)=w^{-k-1}$. Then
it is straightforward to show that we get the correspondence rules
\begin{equation}
k\psi_k \longrightarrow w\partial_w \psi_w \qquad
\psi_{k+1} \longrightarrow \frac{\psi_w}{w} \qquad
\psi_{k-1} \longrightarrow w\psi_w .
\end{equation}
The groundstate wavefunction equation then reads
\begin{equation}
\left[ \epsilon+\frac{m}{2}(\frac{1}{w}-w)-\frac{g}{4}(w^2-\frac{1}{w^2})
\right] \psi_w+\frac{g}{4}w\left[2-w^2-\frac{1}{w^2}\right]
\partial_w \psi_w=0 
\end{equation}
which is a solvable (first order) differential equation.
An acceptable contour for which the series
\begin{equation}
\sum_{k=0}^{\infty} (-1)^k |\psi_k|^2
\end{equation}
(on which depend $G(\epsilon)$) converges
is the interval $]-1,1[$. Solving for $\psi_w$ then
inserting it in the integral representation (\ref{12}), we finally obtain
\begin{equation}
\psi_k=\int_{-1}^{1} dw \frac{w^k}{1-w^2}\left(\frac{1+w}{1-w}\right)^
{\frac{m}{g}}\exp(\frac{\epsilon}{g}\frac{w^2}{w^2-1}) .
\end{equation}
It is noteworthy that this integral has already been obtained in \cite{Luc}.
In this paper, the author treats the problem of the XX spin chain, with disorder in
the couplings, our mass disorder, and bathed in a transverse magnetic field,
playing the role of our mean value $m$ of the random mass field.
His derivation resorts to the characteristic exponent $\Omega(\epsilon)$ of the
localization problem, which can be formally identified to $2\lambda(i\epsilon)$.
This model is known to be ultimately equivalent to the random mass model with zero or
finite mean value of the random mass field.

\begin{figure}[p]
\begin{center}
\begin{tabular}{|c|c|}
\hline
\epsfbox{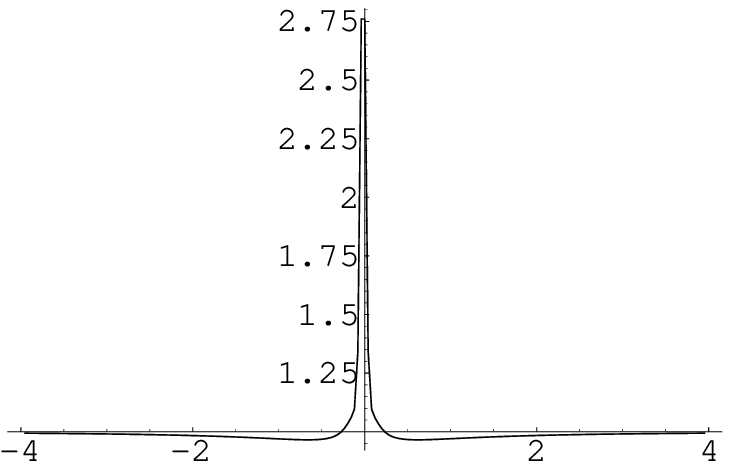} &
\epsfbox{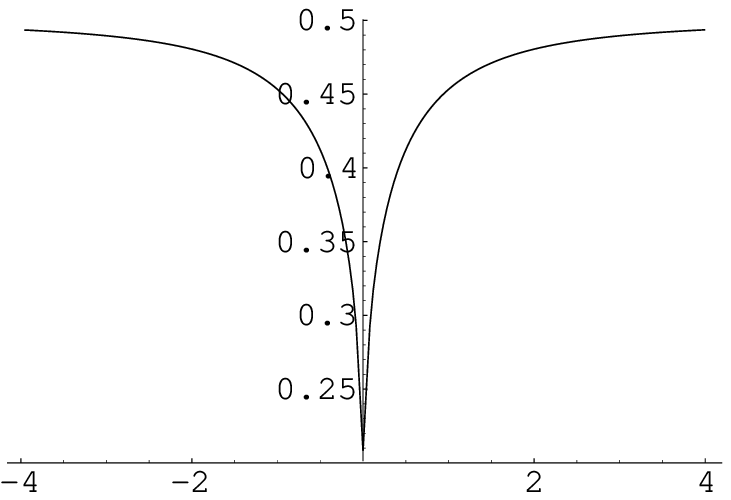} \\
\hline
\end{tabular}
\caption{Density of states and inverse localization length
for the random mass model $g_x=1.$ and $m=0.$}
\label{fig1}
\end{center}
\end{figure}

The latter $\psi_k$ is not normalized. Recall $||\Psi||^2=|\psi_0|^2=1$, so
taking ${\psi_k}/{\psi_0}$ as the wavefunction component normalizes the groundstate.
Thanks to the change of variable $w=\tanh (\frac{\theta}{2})$
 and to some integrations by parts we finally obtain an expression for $\lambda$
\begin{equation}
\lambda=\frac{m}{2}
+\frac{\epsilon}{2}\frac{K_{1-\frac{m}{g}}(\frac{\epsilon}{g})}
{K_{-\frac{m}{g}}(\frac{\epsilon}{g})}
\end{equation}
where $K$ is the modified Bessel function of the second type.
Hence, the resolvent may be put in the form:
\begin{equation}
\label{09}
G(\epsilon)=-i\dpp{}{\epsilon}\left[
\epsilon\frac{K'_{\frac{m}{g}}(\frac{\epsilon}{g})}
{K_{\frac{m}{g}}(\frac{\epsilon}{g})}\right] 
\end{equation}
which is the result obtained in \cite{Com} for $h^2$. The prime denotes derivative with respect
to the argument of the Bessel function.
From this imaginary energy formula, we can derive the integrated density of states
as well of the inverse of the typical localization length.
Indeed the application of Thouless formula to our case yields
\begin{equation}
2\lambda(i\epsilon)=l^{-1}(\epsilon)+i\pi N(\epsilon)
\end{equation}
after analytic continuation on the energy.
The final expressions for the density of states and inverse localization length
can be found in \cite{Com}. They are drawn in figure \ref{fig1}.
They are special cases of the foregoing formulas related to the problem of a random mass
particle in a constant magnetic field (see table \ref{t2}).

	\subsection{Random electric potential model}

	\subsubsection{Massless Dirac particle}

The random electric potential model of a massless particle 
corresponds to the original Hamiltonian
\begin{equation}
h=-i\sigma_z \partial_x+V
\end{equation}
where the electric potential $V$ follows
the Gaussian measure $P(V)=\exp(\frac{1}{2g}\int_{-\infty}^{\infty}dx V^2)$.
The model is trivially soluble because the wavefunction eigenstates can
be explicitly written at all energies
\begin{equation}
\psi_\epsilon (x)=\exp(i\sigma_z\int_0^x [\epsilon-V(u)] du)
\end{equation}
and for any quenched disorder $V$.

\subsubsection{Massive particle}

The random electric potential model of a massive particle 
corresponds to the original Hamiltonian
\begin{equation}
h=-i\sigma_z \partial_x+m\sigma_y+V
\end{equation}
where $m$ is the fixed mass of the particle and $V$, the electric potential follows
the Gaussian measure $P(V)=\exp(\frac{1}{2g}\int_{-\infty}^{\infty}dx V^2)$.
The effective Hamiltonian is $H=\epsilon Z^z-imZ^x+g (Z^z)^2$.
Correspondingly, the groundstate equation is for $k\geq 1$
\begin{equation}
\label{13}
(\epsilon+gk)\psi_k+\frac{m}{2}(\psi_{k+1}-\psi_{k-1})=0
\end{equation}
whereas the groundstate energy is reduced to
\begin{equation}
\label{14}
\lambda=\frac{\epsilon}{2}+\frac{m}{2}\frac{\psi_1}{\psi_0} .
\end{equation}
The careful reader might have noticed that this recurrence relation is the one for
modified Bessel functions. Since we must cautiously take into account boundary conditions,
it is nevertheless safer to use again Laplace method.
For our purpose, we choose once more the Melin kernel, $K(k,w)=w^{-k-1}$.
The groundstate energy then reads
\begin{equation}
\left[ \epsilon+\frac{m}{2}(\frac{1}{w}-w) \right] \psi_w
+gw\partial_w \psi_w=0.
\end{equation}
An acceptable contour is precisely one of those used for the Bessel functions
integral representations.
The contour comes from $-\infty$ slightly below the real axis,
rounds the origin $0$ then goes back to $-\infty$ slightly above the real axis.
Solving for $\psi_w$ then inserting it in the integral representation, we obtain
the non-normalized value of $\psi_k$
\begin{equation}
\psi_k=\int_{-\infty}^{0^+} dw. w^{-k-\frac{\epsilon}{g}-1}
\exp(\frac{m}{2g}(w+\frac{1}{w})).
\end{equation}
We finally get an exact expression for $\lambda$
\begin{equation}
\lambda=\frac{\epsilon}{2}
+\frac{m}{2}\frac{I_{1+\frac{\epsilon}{g}}(\frac{m}{g})}
{I_{\frac{\epsilon}{g}}(\frac{m}{g})}
\end{equation}
where $I$ is the modified Bessel function of the first type.
Thus, after a little algebra,
\begin{equation}
G(\epsilon)=im\dpp{}{\epsilon}\left[
\frac{I'_{\frac{\epsilon}{g}}(\frac{m}{g})}
{I_{\frac{\epsilon}{g}}(\frac{m}{g})}\right].
\end{equation}

In the disorder free model, the mass band $]-m,m[$ is a forbidden energy region.
The density of states and localization length are well-known
and given in table \ref{t1}.

\begin{figure}[p]
\begin{center}
\begin{tabular}{|c|c|}
\hline
\epsfbox{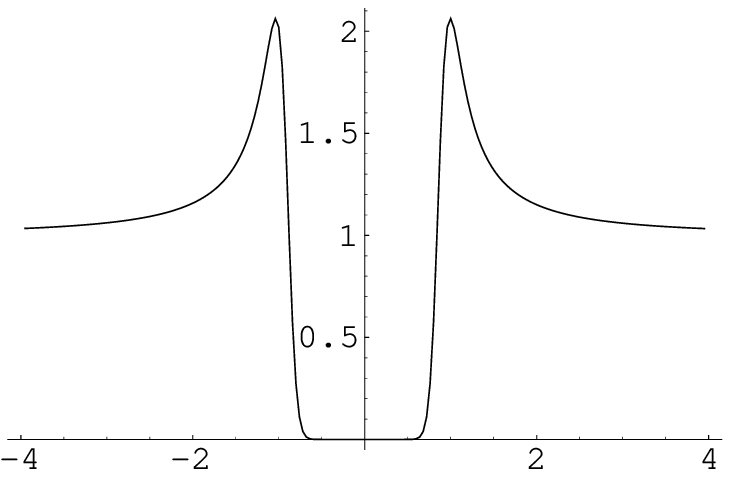} &
\epsfbox{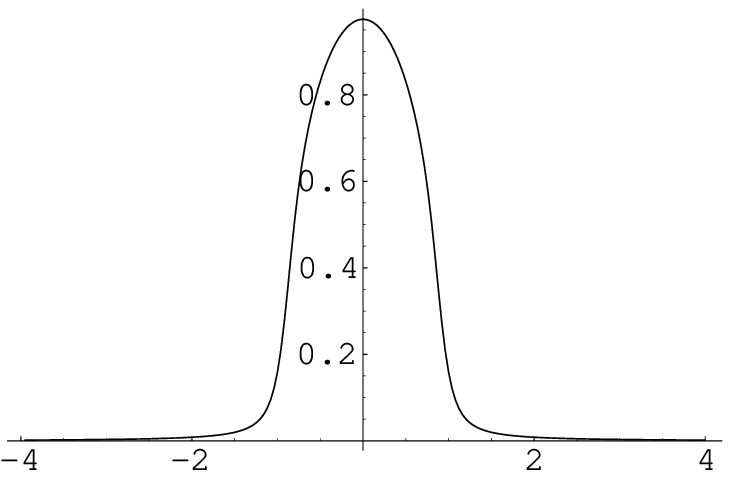} \\
\hline
\end{tabular}
\caption{Density of states and inverse localization length
for the electric potential model $g_z=0.05$ and $m=1.$}
\label{fig2}
\end{center}
\end{figure}

Coming back to the disordered case,
the density of states and inverse localization length can be obtained through
Thouless formula. In that case, it is cumbersome to write definitive expressions
for them because the Bessel functions $I$ bear an {\em imaginary index}.
We have drawn them anyway in figure \ref{fig2}.

When there is disorder, the particle penetrates (statistically speaking)
in the mass band. That is why it is interesting to study the weak disorder limit
in that range of energy.
To derive asymptotic expansion of the density of states and inverse localization
length, we apply the steepest descent method to the
integral representation of the modified Bessel function $I$.
Because the index of the Bessel function is an imaginary index, we cannot
straightforwardly resort to tabulated asymptotics Bessel function expansions.
Surprisingly, this technique only yields real values when expanding $\lambda(i\epsilon)$.
This remark is valid to all orders.
The imaginary part of $\lambda(i\epsilon)$ that accounts for the part
of the density curve in the band $]-m,m[$ is provided by the direction of rapid oscillations
which is transverse to the steepest way in the complex plane.
That is why $\mbox{Im} (\lambda(i\epsilon))$
cannot be accessed through the steepest descent method.

The steepest way expansion is worked out within the $ |\epsilon|< m$ band.
As for the inverse localization length, we find in the limit $g$ goes to $0$
\begin{equation}
l^{-1}(\epsilon)\simeq -gm\dpp{}{m}\left\{ \frac{1}{2}\ln\left(\frac{g}{2\pi m \cos(\alpha)}\right)
+\frac{\epsilon}{g} (\cot (\alpha)+\alpha) \right\} \quad
\mbox{where} \quad \alpha=\arcsin (\frac{\epsilon}{m}).
\end{equation}

As for the integrated density of states, when the energy is smaller enough than the mass, the
imaginary part of the Bessel function becomes crucial.
Fortunately, the imaginary part of the modified Bessel function of imaginary index can be
accessed directly
\begin{equation}
\mbox{Im}(I_{i\frac{\epsilon}{g}}(\frac{m}{g}))=-\frac{\sinh(\frac{\epsilon}{g})}{\pi}.
\frac{\pi}{2i\sin (i\pi\frac{\epsilon}{g})}
\left[ I_{-i\frac{\epsilon}{g}}(\frac{m}{g})-I_{i\frac{\epsilon}{g}}(\frac{m}{g})\right]
=-\frac{\sinh(\frac{\epsilon}{g})}{\pi} K_{i\frac{\epsilon}{g}}(\frac{m}{g})
\end{equation}
which allows an approximate evaluation.
We then find
\begin{equation}
\label{07}
N(\epsilon)\simeq -gm\dpp{}{m}\arctan\left\{ \sinh (\frac{\epsilon}{g})\sqrt{\cos (\alpha)}
\exp(-\frac{m}{g}-\frac{\epsilon^2}{2gm}-
\frac{\epsilon}{g}(\cot (\alpha)+\alpha))\right\}.
\end{equation}

On average, the residual densities of states at $\epsilon=0$ is 
\begin{equation}
\rho(0) \simeq _{g\rightarrow 0} \frac{2m}{\pi g}e^{-2\frac{m}{g}}.
\end{equation}

The average length of energy penetration in the mass band can be extracted from
(\ref{07}) and is of the same order than $\rho(0)$.

	\subsection{Multiple disorder model}

We would like to know what the effect of combining two or three types of disorder is.
We will show that such a combination destroys the $\epsilon=0$ singularity.
To be as simple as possible, we will choose the mean values of those random
fields to be vanishing. Yet the calculations could be generalized to non-zero mean values.
This corresponds to the original Hamiltonian
\begin{equation}
h=-i\sigma_z \partial_x+\Phi \sigma_x+M\sigma_y+V
\end{equation}
where $M$ is the random mass of the particle,
$\Phi$ the random magnetic field and $V$ the electric potential.

The effective Hamiltonian is $H=\epsilon Z^0+g_x (Z^x)^2+g_y (Z^y)^2+g_z (Z^z)^2$.
Correspondingly, the groundstate wavefunction equation reads for $k\geq 1$
\begin{equation}
(\epsilon+(g_z+\frac{g_x+g_y}{2})k)\psi_k+\frac{1}{4}(g_y-g_x)(k-1)\psi_{k-2}+
\frac{1}{4}(g_y-g_x)(k+1)\psi_{k+2}=0
\end{equation}
whereas the groundstate energy is
\begin{equation}
\lambda=\frac{\epsilon}{2}+\frac{g_x+g_y}{4}+\frac{g_y-
g_x}{2}\frac{\psi_2}{\psi_0} .
\end{equation}
Only the even terms contribute. Indeed one can show that the odd terms of the
sequence $\{\psi_k\}$ yield an unacceptable diverging sub-series.

So let us set $\psi_{2k}\rightarrow \psi_k$, define $\delta=g_y-g_x$
and $\sigma=2g_z+g_y+g_x$ then the previous formulas turn to
\begin{eqnarray*}
&&\lambda=\frac{\epsilon}{2}+\frac{g_x+g_y}{4}+\frac{g_y-
g_x}{2}\frac{\psi_1}{\psi_0}, \\
&&(\epsilon+\sigma.k)\psi_k+\frac{\delta}{2}(k-\frac{1}{2})\psi_{k-1}
+\frac{\delta}{2}(k+\frac{1}{2}) \psi_{k+1}=0 .
\end{eqnarray*}
Now introduce the roots of $w^2+2\frac{\sigma}{\delta}w+1=0$:
\begin{equation}
\alpha_{\pm}=-\frac{\sigma}{\delta}\pm \sqrt{\frac{\sigma^2}{\delta^2}-1} 
\end{equation}
and the constant
\begin{equation}
\beta=\frac{2}{\delta(\alpha_+-\alpha_-)}.
\end{equation}
Since $|\frac{\sigma}{\delta}|=|\frac{2g_z+g_x+g_y}{g_y-g_x}|>1$, those roots are 
always real. This is what will guarantee that a critical point at $\epsilon=0$ cannot exist
in presence of at least two types of disorder. To make what follows clearer, we 
choose $\delta<0$ so that the roots $\alpha\pm$ are positive.

\begin{figure}[p]
\begin{center}
\begin{tabular}{|c|c|}
\hline
\epsfbox{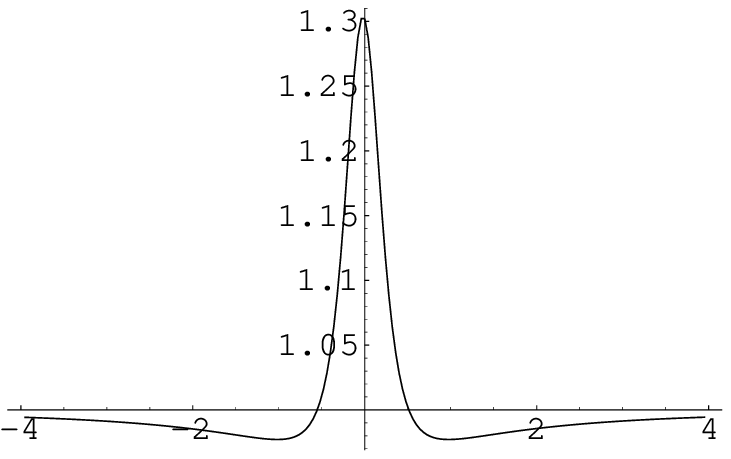} &
\epsfbox{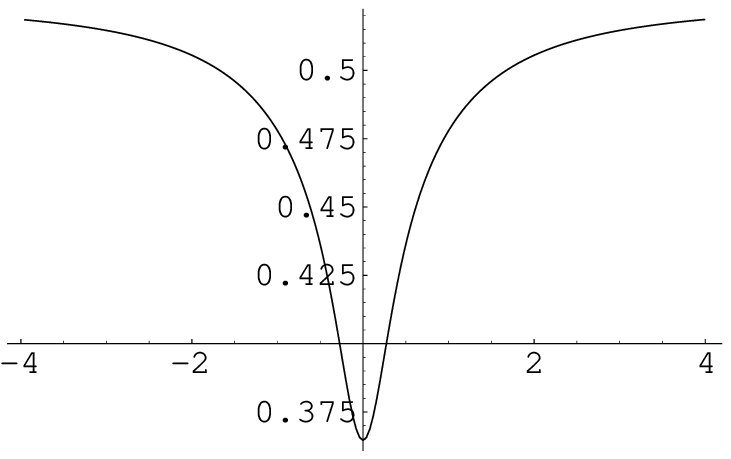} \\
\hline
\end{tabular}
\caption{Density of states and inverse localization length
for the multiple disorder model $g_x=1.$ and $g_y=0.05$}
\label{fig3}
\end{center}
\end{figure}

Using the Laplace method, solving for $\psi_w$, then we find
\begin{eqnarray*}
\psi_k &=& \frac{1}{2i\pi}\oint_{[\alpha_+,\infty[}dw. w^{-k-\frac{1}{2}}.
(w-\alpha_+)^{-\frac{1}{2}-\beta\epsilon}(w-\alpha_-)^{-\frac{1}{2}+\beta\epsilon} \\
&=&\frac{\cos (\pi\beta\epsilon)}{\pi} \int^{\infty}_{\alpha_+} d\alpha. \frac{
(\alpha-\alpha_+)^{-\frac{1}{2}-\beta\epsilon}(\alpha-\alpha_-)^{-
\frac{1}{2}+\beta\epsilon}}
{\alpha^{k+\frac{1}{2}}}
\end{eqnarray*}
so that we obtain $\lambda$ and
\begin{equation}
G(\epsilon)=i\left(1+\frac{g_y-g_x}{2}\dpp{}{\epsilon} \frac{
\int^{\infty}_{\alpha_+} d\alpha. \alpha^{-\frac{3}{2}}.
(\alpha-\alpha_+)^{-\frac{1}{2}-\beta\epsilon}(\alpha-\alpha_-)^{-
\frac{1}{2}+\beta\epsilon}}
{\int^{\infty}_{\alpha_+} d\alpha. \alpha^{-\frac{1}{2}}.
(\alpha-\alpha_+)^{-\frac{1}{2}-\beta\epsilon}(\alpha-\alpha_-)^{-
\frac{1}{2}+\beta\epsilon}}\right).
\end{equation}
Coming back to the original notations, we have:
\begin{equation}
\alpha_{\pm}=-\frac{2g_z+g_x+g_y}{g_y-g_x} \pm
\sqrt{\left(\frac{2g_z+g_x+g_y}{g_y-g_x}\right)^2-1} \qquad \mbox{and} \qquad
\beta=\frac{1}{2\sqrt{(g_z+g_x)(g_z+g_y)}}.
\end{equation}
The typical inverse localization length associated to this imaginary energy resolvent
is finite. The presence of a second type of disorder (or more) makes the
random mass critical point massive.
The density of states and inverse localization length are drawn in figure \ref{fig3}.
	
\subsection{Random mass particle in a constant magnetic field}

\begin{figure}[p]
\begin{center}
\begin{tabular}{|c|c|}
\hline
\epsfbox{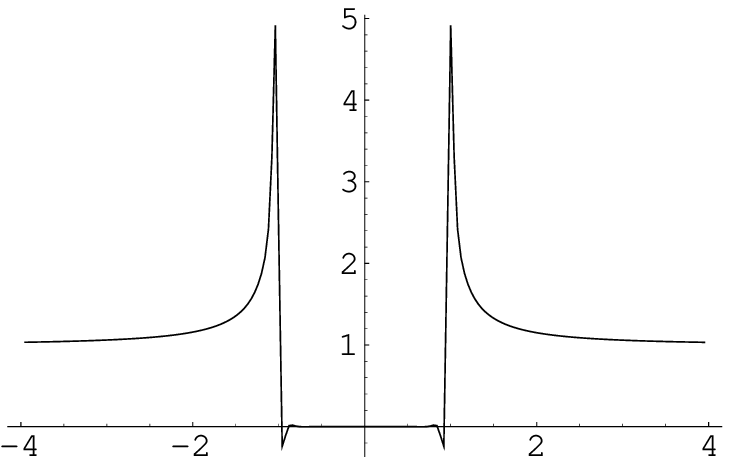} &
\epsfbox{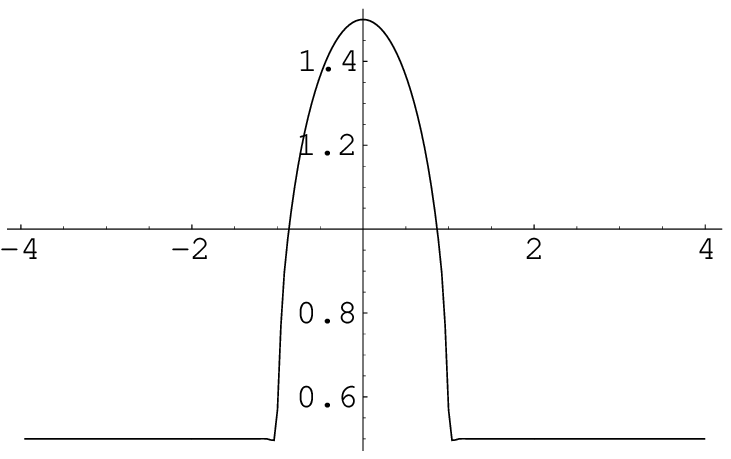} \\
\hline
\end{tabular}
\caption{Density of states and inverse localization length
for the random mass model in constant magnetic field
$g_x=1.$, $m=0.5$ and $\phi=1.$}
\label{fig4}
\end{center}
\end{figure}

This model is equivalent to a Dirac particle with a fixed finite mass
in a random magnetic field. It corresponds to the original Hamiltonian
$ h=-i\sigma_z \partial_x+\phi \sigma_x +M\sigma_y$.
$\phi$ is the homogeneous magnetic-like field and $M$ is the random mass field which
follows the Gaussian measure $P(M)=\exp(-\frac{1}{2g}\int_{-\infty}^{\infty} dx (M-
m)^2)$. $\phi$ can also be interpreted as a uniform staggered magnetic field in the third
direction perturbating an XX random spin chain.
After averaging the effective Hamiltonian one obtains
\begin{equation}
H=\epsilon Z^0+(\phi-im)Z^x+g (Z^x)^2.
\end{equation}
The calculations are of minor difficulty in comparison to the random mass case
provided one uses the non-trivial mapping described in section \ref{sec7}.
So we just give the results:
\begin{equation}
G(\epsilon)=-i\dpp{}{\epsilon}\left[ \sqrt{\epsilon^2+\phi^2}.
\frac{K'_{\frac{m}{g}}(\frac{\sqrt{\epsilon^2+\phi^2}}{g})}
{K_{\frac{m}{g}}(\frac{\sqrt{\epsilon^2+\phi^2}}{g})}\right] .
\end{equation}
Continuing on $\epsilon$ we obtain the expressions for the density of states
and inverse localization length given in table \ref{t2}.
They are also drawn in figure \ref{fig4}.\\

In this section, we have derived imaginary energy resolvent formulas which bring 
about the exact density of states as well as the exact typical localization length in
the cases of a few interesting models. As long as the disorders are Gaussian white noise,
we can calculate those quantities exactly. This is due to the fact that equation (\ref{02})
is linear in $k$. The Laplace method then provides us with a
first order differential equation. That is why an exact explicit expression
can always be obtained.

	\section{Supersymmetric analysis}
\label{sec6}

In \cite{Bal}, the authors make use of the supersymmetric trick to describe
the random mass model in the vicinity of $\epsilon=m=0$.
They obtained equation (\ref{01}) after they identified
the groundstate sector by pre-diagonalizing the effective Hamiltonian.
We will see that this diagonalization is not necessary
and that the sought sector can be algebraically generated as a degenerate $su(1,1)$
grade-star representation. Similarly to the replica version,
we will identify $U(1,1)$ as the
dynamical group of the effective Hamiltonian, preferably to $U(1,1|2)$.
	
The path integral representation of the resolvent can be obtained as in \cite{Bal}.
We recall the formula for the averaged resolvent:
\begin{eqnarray*}
G(\epsilon+i\eta)&=&\langle \mbox{tr}_{\mbox{\tiny spin}} \int \mesb \mesf 
.\psibar_\alpha(x)
\psi_\beta(x) \exp (-\int_{-\infty}^{\infty}dx
\xi^*(x) \langle x|-ih+(-i\epsilon+\eta)|x \rangle \xi(x) \\
&&-\int_{-\infty}^{\infty}dx\psibar(x) \langle x|-ih+(-i\epsilon+\eta)|x
\rangle \psi(x)) \rangle .
\end{eqnarray*}
	
As in the replica case, we can now rename $-i\epsilon+\eta$ as $\epsilon$ and make it
strictly positive. When all calculations are done, we will continue on
$\epsilon$ to re-establish it to the imaginary axis.
Similarly to the replica case, the previous path integral formula is ill defined,
at least as far as the Bosons are concerned. To see it, we follow the
Feynman procedure and write the classical Hamiltonian associated to the
previous path-integral. We will then quantize it.
The momenta for the spinorial fields $\xi$ and $\psi$ are
\begin{equation}
\Pi_\psi=\frac{\delta {\cal L}}{\delta \dot{\psi}}=-\psibar \sigma_z
\quad \mbox{ and } \quad
\pi_\xi=\frac{\delta {\cal L}}{\delta \dot{\xi}}=- \xi^*\sigma_z
\end{equation}
where the dots indicate derivative with respect to $x$.
Here the Fermionic derivative is a right derivative.
The Euclidean coherent state path integral formalism yields the
classical Hamiltonian
\begin{equation}
H_c=\int dx \Pi_\psi.\partial_x \psi+
{\cal L}(\Pi_\psi,\psi)+\Pi_\xi.\partial_x \xi+ {\cal L}(\Pi_\xi,\xi)
\end{equation}
before any ensemble averaging.

To quantize this Hamiltonian, we impose the following
canonical commutator:
\begin{equation}
\{\psi(y), \Pi_\psi(x)\}=\delta(x-y) \qquad
[\xi(y), \Pi_\xi(x) ]=\delta(x-y) .
\end{equation}
This leads to a $\epsilon.\Pi_{\xi}\sigma_z \xi$ term which makes the
quantum Hamiltonian unbounded, due to the minus sign appearing in the product
of the spin down component operators.
To quantize the Hamiltonian safely we have the freedom
to make any canonical transformation on the spin component operators.
For the Dirac Fermions the correct procedure is to fill the Dirac sea
upon making the particle hole transformation:
\begin{equation}
\Pi_\psi=(\psi^+_\uparrow,\psi_\downarrow) \qquad
\psi=(\psi_\uparrow,\psi^+_\downarrow) .
\end{equation}
However, as we already noticed in the replica case, such a procedure
is inappropriate for the Bosons. As before we need a stronger transformation.
We then set
\begin{equation}
\Pi_\xi=(\xi^+_\uparrow,-\xi_\downarrow) \qquad
\xi=(\xi_\uparrow,\xi^+_\downarrow) .
\end{equation}  
The minus sign heals the unboundedness of the spectrum and
the additional particle-hole transformation restores the commutation
relation destroyed by the inversion of the spin down component operators.
Again, as a result, the spinor momentum
$\Pi_{\xi}$ is not the Hermitian conjugate of $\xi$.
The internal dynamical symmetry of the
supersymmetric Hamiltonian is not any more $U(2|2)$ but $U(2|1,1)$. 

In order to go beyond this dynamical symmetry of the Hamiltonian, let us define
the following Hermitian spin operators
\begin{eqnarray*}
&{\cal J}^x=\frac{i}{2}(\bspu\bspd-\bsmu\bsmd)
+\frac{i}{2}(\fspu\fspd-\fsmu\fsmd)  \qquad
{\cal J}^y=\frac{1}{2}(\bspu\bspd+\bsmu\bsmd)
+\frac{i}{2}(\fspu\fspd+\fsmu\fsmd)  & \\
&{\cal J}^z= \frac{1}{2}(\bspu\bsmu+\bspd\bsmd)
+\frac{1}{2}(\fspu\fsmu+\fspd\fsmd) .&
\end{eqnarray*}
Like in the replica case, we point out that
the Hamiltonian does only depend on those sum operators.
Then we define the corresponding ladder operators 
\begin{equation}
{\cal J}^+={\cal J}^y-i{\cal J}^x
\qquad {\cal J}^-={\cal J}^y+i{\cal J}^x
 \qquad {\cal J}^0={\cal J}^z
\end{equation}
which obey $su(1,1)$ commutation rules:
\begin{equation}
[{\cal J}^+,{\cal J}^-]=-2{\cal J}^0 \qquad [{\cal J}^0,
{\cal J}^-]={\cal J}^+ \qquad [{\cal J}^0,{\cal J}^-]=-{\cal J}^-.
\end{equation}
However, as such, this statement is incorrect because we cannot avoid
$({\cal J}^+)^\dagger\neq {\cal J}^-$. By computing the commutation relations
of the spin operators, we merely proved they generate a $sl(2,C)$ algebra but
did not stipulate its real form.
The reason of that difference is that although the Bosonic part of ${\cal J}^+$
represents a Bosonic $su(1,1)$ spin, the Fermionic part represents a $su(2)$ spin.
	
Now we shall show how to "transmute" this last spin into a $su(1,1)$ one.
Such a transformation is conceivable because the Fock space is a grade space.
Hence two types of star representations are permitted. We shall resort to the
unusual one: the grade-star representation. It is characterized by the properties
of the adjoint operation \cite{Sch} on the Lie algebra operators:
\begin{equation}
(aP+bQ)^\dagger=\overline{a}P^\dagger+\overline{b}Q^\dagger \qquad
[P,Q]^\dagger=(-1)^{\deg P.\deg Q}[P^\dagger,Q^\dagger]  \qquad
(Q^+)^\dagger=(-1)^{\deg Q}Q.
\end{equation}
$a,b$ are c-number and $P,Q$ are operators.
$\deg P$ takes the value $0$ or $1$ whether $P$ is even or odd.

In order to satisfy such identities, we must change our Fock space
scalar product. It was previously implicitly defined
by the creation/annihilation operators and how they act on the vacuum.
Now assume that instead of the usual definition of conjugation for Fermionic operators
\begin{equation}
(\fsmu)^\dagger=\fspu \qquad (\fspu)^\dagger=\fsmu \qquad \mbox{and} \qquad
(\fsmd)^\dagger=\fspd \qquad (\fspd)^\dagger=\fsmd
\end{equation}
we prefer
\begin{equation}
(\fsmu)^\dagger=\fspu \qquad (\fspu)^\dagger=-\fsmu \qquad \mbox{and} \qquad
(\fsmd)^\dagger=\fspd \qquad (\fspd)^\dagger=-\fsmd
\end{equation}
which are coherent assumptions provided we accept to redefine the scalar product
in the Fermionic sector of the Fock space.
One can check the compatibility of this new {\em scalar product} with the grade-star
adjoint operation through
\begin{equation}
\langle P^\dagger x | y\rangle=(-1)^{\deg P.\deg x}\langle x | P y\rangle 
\end{equation}
for any homogeneous operator $P$ and homogeneous vectors $x$ and $y$.
Yet, this {\em scalar product} is not any more positive definite
(that is why in principle we should not call it any more scalar product).
For example if $| \Omega \rangle$ is the Fermionic vacuum of norm unity, then
$|| \fspu | \Omega \rangle ||^2=1$ and $|| \fspu\fspd | \Omega \rangle ||^2=-1$.
More generally a state with $4n$ or $4n+1$ Fermions has norm $1$, whereas
a state with $4n+2$ or $4n+3$ Fermions has norm $-1$.

A consequence of this change is that we recover $({\cal J^+})^\dagger={\cal J}^-$.
Another related consequence is that the effective Hamiltonian in real time and when
$\eta=0$ (physical limit) is a true Hermitian operator, provided we accept the 
degenerate scalar product.
As a result the $\vec{\cal J}$ spin operators now thoroughly obey the
constraints of a $su(1,1)$ algebra.
This statement implies that the dynamical algebra can be reduced from $u(1,1|2)$
to $u(1,1)$. Such a strong simplification is analogous to what occurs
in the replica derivation.

A practical consequence is that once we identified the (grade) star representation
of the dynamical symmetry of the Hamiltonian, we can easily generate its main
sectors. For example the groundstate sector is generated by the states
$|k\rangle=\frac{({\cal J}^+)^k}{k!} |\Omega \rangle$, where now $|\Omega \rangle$ 
stands for both the Bosonic and Fermionic vacuum.
Within our formalism we can now write them explicitly in terms of creation operators:
\begin{equation}
|k\rangle=\frac{({\cal J}^+)^k}{k!} |\Omega \rangle=
\frac{1}{k!}(\bspu\bspd+\fspu\fspd)^k |\Omega \rangle=
\frac{1}{k!}((\bspu\bspd)^k+k(\bspu\bspd)^{k-1}\fspu\fspd)^k |\Omega \rangle
\end{equation}
which are precisely the states obtained in \cite{Bal}.

The left eigenstates can be mapped from the vector dual space to the usual
space by means of the unitary operator $R=\exp(i\pi {\cal J}^z)$ \footnote{
It must not be confused with the Bosonic operator used in \cite{Bal}.},
which is a supersymmetric operator.
The action of the general Hamiltonian
\begin{equation}
H=\epsilon {\cal J}^0-im{\cal J}^x+i\phi {\cal J}^y
+g_x ({\cal J}^x)^2+g_y ({\cal J}^y)^2+g_z ({\cal J}^z)^2
\end{equation}
on the expanded groundstate $|\Psi\rangle=\sum_{k=0}^{\infty} \zeta_k |k \rangle$
produces 0-order equation (\ref{02}).
Contrary to the replica analog, the supersymmetric method only leads to the sum
formula (\ref{05}) for the resolvent.
Indeed, using the mapping from path integral formulation to quantum mechanics,
we may only write the resolvent as
\begin{equation}
G(\epsilon)=-2i
\langle \Psi|_l S^0 |\Psi\rangle_r.
\end{equation}
where $S^0$ is the fermionic part
of the generator ${\cal J}^0$. It is the analog of (\ref{11})
and leads to formula (\ref{05}).

\section{Symmetries of disorders}
\label{sec7}

In this section, we discuss some mappings that may exist between Hamiltonians
with different disorders.
Such question is of most interest in the 2D disordered Dirac Fermions
model. Indeed it has been argued that the renormalization group flow of the random 
mass model can be, in some sense, mapped to the random electric potential model \cite{Lud}.
In this one-dimensional case, not only does it exist at least one
non-trivial mapping between disorders, but it can also be made explicit. Besides
there exists a more obvious mapping between magnetic and mass disorder which
deserves quoting.

The geometric group with double covering group $SU(1,1)$ is the Lorentz group
$SO(2,1)$. So we may imagine the 3D space of disorder coupling strengths
in the following way.
The time (Oz-)axis measures the potential electric coupling strength whereas
the Ox and Oy axis measure the mass field and magnetic field strengths.
The $SO(2,1)-$ invariant manifolds are the hyperbolo\"\i ds of axis Oz.
${\cal J}^x$, ${\cal J}^y$ and ${\cal J}^z$ are the generating operators
of the Lie algebra.
Precisely ${\cal J}^z$ generates the Euclidean rotation around axis Oz,
whereas ${\cal J}^x$ and ${\cal J}^y$ generate boosts of axis Ox and Oy.
As a consequence, we can write the action of a half-turn rotation around the
time axis on the spin operators:
\begin{equation}
e^{-i\pi {\cal J}^z}{\cal J}^x e^{i\pi {\cal J}^z}={\cal J}^y \qquad
e^{-i\pi {\cal J}^z}{\cal J}^y e^{i\pi {\cal J}^z}=-{\cal J}^x.
\end{equation}
This rotation operator establishes a unitary mapping between physics of the random
magnetic field and physics of the random mass field.

More interesting is the remark that no unitary transformation can map a mass disorder
(or a magnetic disorder) to a potential electric disorder. Indeed no unitary boost
is able to map a spatial axis to the time axis.
Let us now consider the non-unitary, yet non-singular, operator
$R=\exp(\frac{\pi}{2} {\cal J}^y)$
which is a boost of axis Oy and imaginary angle $i\frac{\pi}{2}$.
We will show that it maps a random mass problem to an electric potential model.
In particular we have these actions on the spin operators:
\begin{equation}
e^{\frac{\pi}{2} {\cal J}^y}{\cal J}^z e^{-\frac{\pi}{2} {\cal J}^y}=i{\cal J}^x
 \qquad e^{\frac{\pi}{2} {\cal J}^y}{\cal J}^x e^{-\frac{\pi}{2} {\cal J}^y}=
i{\cal J}^z.
\end{equation}

Let us start reasoning from the effective Hamiltonian
$H=\epsilon {\cal J}^z-im{\cal J}^x+g({\cal J}^x)^2$ of the random mass model.
Then we make $R$ act on $H$:
\begin{equation}
R H R^{-1}=i\epsilon {\cal J}^x+m{\cal J}^z-g({\cal J}^z)^2.
\end{equation}
Upon making the replacements
\begin{equation}
\epsilon\longrightarrow m \qquad m\longrightarrow -\epsilon \qquad
\mbox{and} \qquad g \rightarrow -g
\end{equation}
we turn the random electric potential problem of section \ref{sec5} into the model
describe by $R H R^{-1}$.

However being non-unitary the transformation $R$ is not trivial.
Above all we cannot use scalar-product-dependent properties of the Hamiltonians
since we lost unitarity.
That is why in such circumstances, we must consider our Fock space not as an 
Hilbert space but as a mere vector space.

Yet we notice that the kernels of operators are covariant objects, i.e.
$\mbox{Ker}(RHR^{-1})=R( \mbox{Ker}(H))$ because $R$ is non-singular.
Nevertheless under suppression of the scalar product, the kernel of $H$ has
enlarged from a one-dimensional subspace to a two-dimensional one, because
any normalization condition has been dropped.
Let us first describe $\mbox{Ker}(\epsilon {\cal J}^z-im{\cal J}^x
+g({\cal J}^z)^2)$. We already know that the wavefunction
$I_{k+\frac{\epsilon}{g}}(\frac{m}{g})$ belongs to the kernel.
A second independent solution is $(-1)^k K_{k+\frac{\epsilon}{g}}(\frac{m}{g})$.
To check this last statement, change variable in the recurrence equation (\ref{13})
from $\psi_k$ to $(-1)^k\psi_k$, and observe it
became the recurrence relation for the modified Bessel function of second type.
So we conclude:
\begin{equation}
\mbox{Ker}(R H R^{-1})=\mbox{Vec}\{ I_{k-\frac{m}{g}}(\frac{\epsilon}{g}),
(-1)^k K_{k-\frac{m}{g}}(\frac{\epsilon}{g})\}
\end{equation}
Our task is then to find out in a two-dimensional kernel what the {\em physical}
image of our original groundstate is. 
We must select the right combination in the 2-dimensional space which stands
for the groundstate wavefunction of our primary problem to perform such a selection.
This combination does not depend on the values of the parameters of the Hamiltonian because $R$ 
acts linearly on the states.

To proceed we can use the alternative method of the replica and
view $n$ as a deformation parameter. Then the groundstate energy of the replicated
Hamiltonian $\lambda(n)=\lambda n+O(n^2)$ is characterized by
\begin{equation}
\label{03}
\lambda=\frac{m}{2}-\frac{\epsilon}{2}\frac{\psi_1}{\psi_0}.
\end{equation}
In the supersymmetric version of the problem, the selection
rule of the groundstate is the convergence of the series defining
the correlation functions. It applies in particular to the resolvent.
In the replica case, the analogous selection rule is
the sign of $\lambda$ which states whether the quantum theory is correctly defined
(bounded or not).
Since we had to forget about the scalar product, it is cumbersome to evaluate the
series defining the resolvent in the original problem, whereas it is much simpler to 
check the sign of $\lambda$ for any reasonable values of parameters $\epsilon$, $m$, and 
$g$.

So we can now apply our selection rule. Essentially because a minus sign appears
in formula (\ref{03}) in comparison to the original formula (\ref{14}),
the only acceptable combination in $\mbox{Ker}(R H R^{-1})$ is the wavefunction
$\psi_k=(-1)^k K_{k-\frac{m}{g}}(\frac{\epsilon}{g})$.
Finally, we conclude 
\begin{equation} 
G(\epsilon)=2i\dpp{}{\epsilon}\lambda=-i\dpp{}{\epsilon} \left[
\epsilon\frac{K'_{\frac{m}{g}}(\frac{\epsilon}{g})}
{K_{\frac{m}{g}}(\frac{\epsilon}{g})}\right]
\end{equation}
and recover result (\ref{09}).
Practically, this way to obtain (\ref{09}) saves many calculations.
In particular we transform a recursion relation of order 5
to another one of order 3, easier to solve.

\section{Conclusion}
\label{sec8}
	
We have calculated the exact density of states and the typical localization length
in a few interesting cases which are: the random electric potential model,
the multiple disorder model, the random mass model in presence of a constant
magnetic field. Doing so, we used efficient algebraic methods to pinch
the desired sectors of the Hamiltonian. For such quantities, the replica treatment
is more straightforward thanks to the use of the replica index $n$ viewed as
a deformation parameter. The dynamical algebra of the effective Hamiltonian
is $u(1,1)$ and allows to generate easily the groundstate sector from which
stem the thermodynamical properties of the models.
The vector space of this groundstate sector is identified
as a supplementary representation of the dynamical algebra, labelled by $n$.
An analog derivation is done in the supersymmetric formalism.

We have also pointed out the existence of mappings between Hamiltonians describing
different disordered models. In particular we mapped a random mass problem to
random scalar potential problem by means of a non-unitary transformation. 

Those calculations can only be partially generalized to the calculation of the
two-point Green function. The algebraic simplifications in the replica and 
supersymmetric cases we took advantage of in pointing out the dynamical algebra are still
valid, as well as the use of the Laplace method. Yet the differential
equations that come about are invariably the confluent equations
of a six elementary singularities second order differential equation.
Unfortunately this forbids exact calculations. Besides approximations
can only be done safely in the case where there is a critical point
and its simplifying properties, that is in the already well-know random mass case.

\begin{acknowledgements}
It is a pleasure to thank C.Monthus, V.Pasquier, A.Comtet, M.Berg\`ere for useful discussions
and T.Jolicoeur for careful reading of the manuscript.    
\end{acknowledgements}

\newpage

\begin{table}
\begin{tabular}{|c||c|c|}
$ \epsilon $ & $\epsilon<m$ & $m<\epsilon $ \\
\hline \hline
$\rho(\epsilon)$ & $0$ & $\frac{1}{\pi}\frac{\epsilon}{\sqrt{\epsilon^2-m^2}}$\\
\hline
$\l(\epsilon)$ & $\frac{1}{\sqrt{1-\frac{\epsilon^2}{m^2}}}$ & $\infty$ \\
\end{tabular}
\caption{Density of states and localization length of free massive Dirac particle.}
\label{t1}
\end{table}

\begin{table}
\begin{tabular}{|c||c|c|}
$ \epsilon^2 $ & $ \epsilon^2<\phi^2 $ & $ \phi^2 <\epsilon^2  $ \\
\hline \hline
$ N(\epsilon)$ & $0$ & $ \frac{2g}{\pi^2}.\frac{1}{M^2_{\frac{m}{g}}
(\frac{\sqrt{\epsilon^2-\phi^2}}{g})} $ \\
\hline
$ \l^{-1}(\epsilon)$ & $-\dpp{}{\epsilon}\left[ \sqrt{\phi^2-\epsilon^2}.
\frac{K'_{\frac{m}{g}}(\frac{\sqrt{\phi^2-\epsilon^2}}{g})}
{K_{\frac{m}{g}}(\frac{\sqrt{\phi^2-\epsilon^2}}{g})}\right] $ &
$ -\frac{\epsilon^2-\phi^2}{\epsilon}\dpp{}{\epsilon}
\ln \left[ M^2_{\frac{m}{g}}(\frac{\sqrt{\epsilon^2-\phi^2}}{g}) \right] $ \\
\end{tabular}
\caption{Average integrated density of states and inverse localization length of
a random mass Dirac particle in a constant magnetic field. $M^2_\mu (z)=J^2_\mu(z)+N^2_\mu (z)$,
and $J_\mu$ and $N_\mu$ are respectively the Bessel functions of first and second type.}
\label{t2}
\end{table}


\end{document}